\documentclass[aps,
               prl,
               twocolumn,
               showpacs,
               amsmath,
               amssymb,
               superscriptaddress,
               reprint,
               10pt
              ]{revtex4-1}

\usepackage[hyperindex,colorlinks,breaklinks,allcolors=blue]{hyperref}
\usepackage{txfonts}

\usepackage{amsmath, amssymb}

\usepackage{pstricks}
\usepackage{graphicx}
\usepackage{tikz}
\usepackage[utf8x]{inputenc}
\usepackage{color}

\usepackage{pstool}
\usepackage{pgf}
\usepackage{import}

\usetikzlibrary{snakes}
\usetikzlibrary{arrows}

\usepackage{tabularx}
\usepackage{multirow}
\usepackage{array}
\newcolumntype{L}[1]{>{\raggedright\let\newline\\\arraybackslash\hspace{0pt}}m{#1}}
\newcolumntype{C}[1]{>{\centering\let\newline\\\arraybackslash\hspace{0pt}}m{#1}}
\newcolumntype{R}[1]{>{\raggedleft\let\newline\\\arraybackslash\hspace{0pt}}m{#1}}

\newcommand{\eq}[1]{(\ref{eq:#1})}

\newcommand{\Fig}[1]{Fig.~\ref{fig:#1}}
\newcommand{\fig}[1]{\ref{fig:#1}}

\makeatletter
\let\cat@comma@active\@empty
\makeatother

\begin{document}

%==============================================================================
%==============================================================================
\title{Prescaling in a far-from-equilibrium Bose gas}

\author{Christian-Marcel Schmied}
\email{christian-marcel.schmied@kip.uni-heidelberg.de}
\affiliation{Kirchhoff-Institut f\"ur Physik, 
             Ruprecht-Karls-Universit\"at Heidelberg,
             Im Neuenheimer Feld 227, 
             69120 Heidelberg, Germany}
\affiliation{Dodd-Walls Centre for Photonic and Quantum Technologies,
             Department of Physics, 
             University of Otago, 
             Dunedin 9016, 
             New Zealand}

\author{Aleksandr N. Mikheev}
\email{aleksandr.mikheev@kip.uni-heidelberg.de}
\affiliation{Kirchhoff-Institut f\"ur Physik, 
             Ruprecht-Karls-Universit\"at Heidelberg,
             Im Neuenheimer Feld 227, 
             69120 Heidelberg, Germany}

\author{Thomas~Gasenzer}
\email{t.gasenzer@uni-heidelberg.de}
\affiliation{Kirchhoff-Institut f\"ur Physik, 
             Ruprecht-Karls-Universit\"at Heidelberg,
             Im Neuenheimer Feld 227, 
             69120 Heidelberg, Germany}

\date{\today}

%==============================================================================
%==============================================================================
\begin{abstract}
Non-equilibrium conditions give rise to classes of universally evolving configurations of quantum-many body systems at non-thermal fixed points.
While the fixed point and thus full scaling in space and time is generically reached at very long evolution times, we propose that systems can show prescaling much earlier in time, in particular, on experimentally accessible time scales.
During the prescaling evolution, some well-measurable properties of spatial correlations already scale with the universal exponents of the fixed point while others still show scaling violations.
Prescaling is characterized by the evolution obeying conservation laws associated with the remaining symmetry which also defines the universality class of the asymptotically reached non-thermal fixed point. 
Here we consider $N=3$ species of spatially uniform three-dimensional Bose gases, with identical inter- and intra-species interactions.
During prescaling, the full $U(N)$ symmetry of the model is broken to $U(N-1)$ while the conserved transport, reflecting explicit and emerging symmetries, leads to the buildup of rescaling quasicondensate distributions.
\end{abstract}

% insert suggested PACS numbers in braces on next line
%\pacs{03.75.Ss, 05.30.Fk, 05.70.Ln, 11.15.Pg, 51.10.+y, 67.10.Jn}

\maketitle

%==============================================================================
%==============================================================================
%

Far from equilibrium, comparatively little is known about the possibilities nature reserves for the structure and states of quantum many-body systems.
Much progress has been made recently in the context of prethermalization \cite{Gring2011a,Berges:2004ce}, generalized Gibbs ensembles \cite{Langen2015b.Science348.207,Jaynes1957a},
many-body localization \cite{Schreiber2015a.Science349.842}, critical and prethermal dynamics \cite{Braun2014a.arXiv1403.7199B,Nicklas:2015gwa,Navon2015a.Science.347.167N,Eigen2018a.arXiv180509802E}, decoherence and revivals \cite{Rauer2017a.arXiv170508231R.Science360.307},  and (wave) turbulence \cite{Zakharov1992a,Navon2016a.Nature.539.72,Johnstone2018a.arXiv180106952}.

Quantum systems quenched far from equilibrium can show relaxation behavior distinctly different from what is known in classical statistics. 
In particular, a system can approach a non-thermal fixed point \cite{Berges:2008wm} exhibiting universal scaling in time and space \cite{Orioli:2015dxa,Prufer:2018hto,Erne:2018gmz}.
Universal behavior has been predicted to occur in various different systems ranging from the post-inflationary early universe \cite{Kofman:1994rk,Micha:2002ey}, via the dynamics of quark-gluon matter created in heavy-ion collisions \cite{Baier:2000sb,Berges:2013eia}, to the evolution of dilute quantum gases starting from a far-from-equilibrium initial state  \cite{Orioli:2015dxa,Berges:2015kfa,Chantesana:2018qsb.PhysRevA.99.043620,Berges:2010ez,Nowak:2010tm,Nowak:2011sk,Schole:2012kt,Karl2017b.NJP19.093014}.
The concept of non-thermal fixed points
paves the way to a unifying description of universal dynamics.
It remains, though, an unresolved question how in general quantum many-body systems evolve from a given initial state to such  a fixed point. 
In this work, we propose \emph{prescaling} as a generic feature of that evolution.

%===============================================
\begin{figure*}
\includegraphics[width=0.9\textwidth]{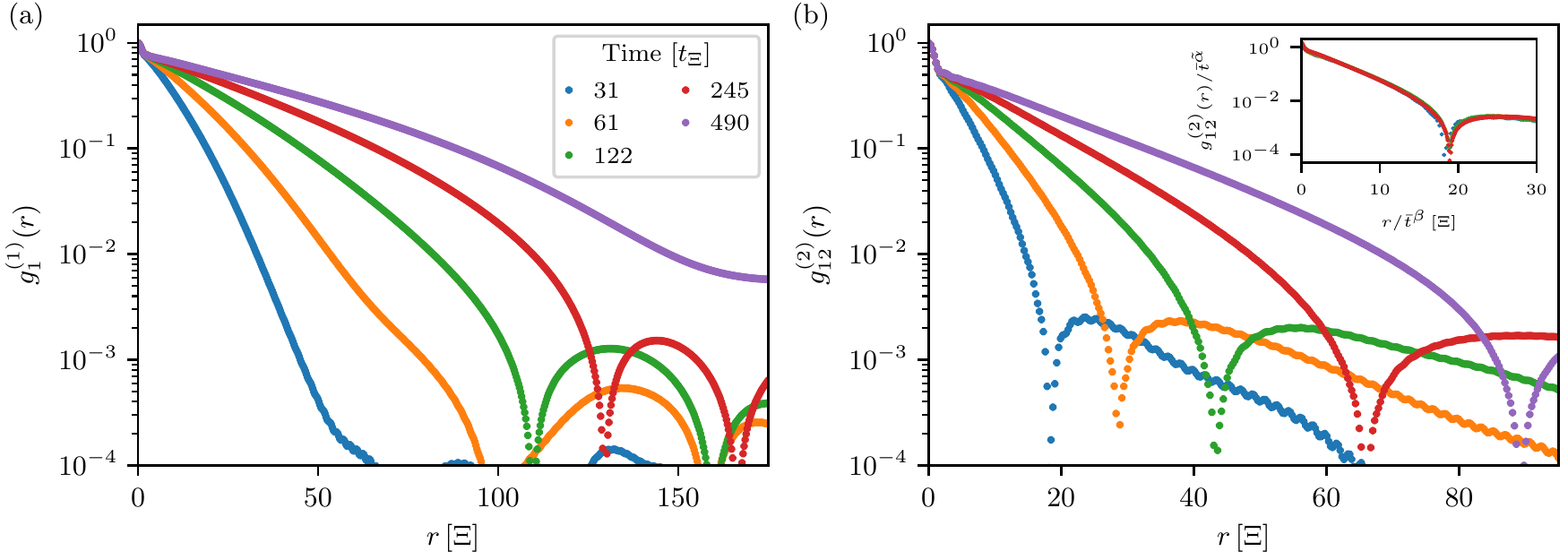}
\caption{%
(a) Time evolution of the first-order coherence function $g^{(1)}_{1}(r)=g^{(1)}_{1}(\mathbf{r},t) = \langle\Phi_{1}^{\dag}(\mathbf{x}+\mathbf{r},t)\Phi_{1}(\mathbf{x},t)\rangle$ at five different times (colored dots).
The shape of the correlation function is reminiscent of an exponential with a multiplicative oscillatory contribution.
It clearly exhibits violations of universal scaling at larger distances, which become weaker in time but still prevail even at long evolution times.
At the latest time shown finite-size effects appear. 
(b) Corresponding second-order coherence function measuring the spatial fluctuations of the relative phases between components 1 and 2, $g_{12}^{(2)} (r,t) = \langle\Phi_{1}^{\dag}(\mathbf{x}+\mathbf{r},t)\Phi_{2}(\mathbf{x}+\mathbf{r},t)\Phi_{2}^{\dag}(\mathbf{x},t)\Phi_{1}(\mathbf{x},t)\rangle$ for the same evolution times as in (a) (colored dots).
The inset shows the rescaled coherence function $\bar{t}^{-\tilde\alpha}g_{12}^{(2)}(\bar{t}^{-\beta}r,t_\mathrm{ref})$, with $\beta = 0.6$, $\tilde{\alpha} = -0.2$, and $\bar{t} = t/t_{\mathrm{ref}}$, with reference time $t_\mathrm{ref}= 31 \, t_\Xi$.
The collapse of the data onto a single function, especially at short distances where $g_{12}^{(2)} (r,t)\gtrsim10^{-2}$, indicates that violations of scaling are considerably weaker than for $g^{(1)}_{1}$.  Time $t$ is measured in units of  $t_{\Xi} = 2\pi[g \rho^{(0)}]^{-1}$, distances $r$ in units of the healing length scale $\Xi = [2mg\rho^{(0)}]^{-1/2}$. 
\label{fig:RelativePhase12}}
\end{figure*}
%===============================================

Universal scaling dynamics associated with a non-thermal fixed point is characterized by scaling evolution of correlation functions.
For example, the occupation number $n_{a}(\mathbf{k},t)=\langle\Phi_{a}^{\dag}(\mathbf{k},t)\Phi_{a}(\mathbf{k},t)\rangle$ of an ($N$-component) Bose field $\Phi_{a}(\mathbf{k},t)$, \emph{at} the fixed point, evolves in a self-similar manner according to
\begin{equation}
n_{a}(\mathbf{k},t) = (t/t_\mathrm{ref})^{\alpha}f_{\mathrm{S},a}([t/t_\mathrm{ref}]^{\beta}\mathbf{k})\,,
 \label{eq:NTFPscaling}
\end{equation}
with universal scaling function $f_{\mathrm{S},a}(\mathbf{k})=n_{a}(\mathbf{k},t_\mathrm{ref})$ depending on a single $d$-dimensional variable  only, scaling exponents $\alpha$, $\beta$, and some reference time $t_\mathrm{ref}$ within the temporal scaling regime  \cite{Orioli:2015dxa}.
In particular, the scaling exponent $\beta$ defines the time evolution of a single characteristic length scale $L_\Lambda (t) \sim t^{\,\beta}$.
Strictly speaking, the fixed point itself is reached only in a certain scaling limit, such as, for $\beta>0$, at asymptotic times and infinite volume.
However, the question arises how the scaling limit is reached and to what extent and when scaling is already seen at finite times.

In equilibrium, fixed points of renormalization-group flows describe correlations at a continuous, e.g.~second-order phase transition.
They correspond to a pure rescaling of the correlations, in momentum or position space, under the change of the flow parameter such as a scale beyond which fluctuations are averaged over.
In the context of critical phenomena as well as fundamental particle physics, renormalization flows are known which are first attracted to a partial fixed point \cite{Wetterich1981a}.
In such situations, still away from the actual fixed point, scaling violations can occur for some quantities while others already show scaling  and the further flow be strongly constrained by a symmetry the system is subject to.

Motivated by the general concept of partial fixed points \cite{Aarts2000a.PhysRevD.63.025012}, we propose the existence of \emph{prescaling} \footnote{C. Wetterich, private communication.}.
This means that certain correlation functions, already at comparatively early times and within a limited range of distances scale with the universal exponents predicted for the fixed point which itself is reached only much later in time and in a finite-size system may not be reached at all.
During the stage of prescaling, (weak) scaling violations occur in correlations at distances outside this range. 
Such violations only slowly vanish as time evolves.
In analogy to the case of partial fixed points, we expect the underlying symmetries of the system to play a key role for the realization of prescaling.
While part of the symmetries can be broken, symmetries reflecting the conservation laws associated with the non-thermal fixed point remain intact during prescaling.     

To reveal the existence of prescaling we employ an isolated, ($N=3$)-component dilute Bose gas in $d=3$ spatial dimensions, quenched far out of equilibrium.
Numerically solving the field equations of motion within a semi-classical Truncated-Wigner approach we find that, during the approach of a non-thermal fixed point, the system prescales.
The phenomenon becomes visible in the short-distance properties of correlation functions that measure, for example, the spatial coherence of the local phase-angle differences between different components. 
We  emphasize that scaling violations affect not only the scaling exponents but in particular also the shape of the scaling functions.

%==============================================================================
The spatially uniform Bose gases consist of identical particles distinguished only by a single property such as the hyperfine magnetic quantum numbers of the atoms forming the gas.
The system in three spatial dimensions is described by a $U(3)$ symmetric Gross-Pitaevskii (GP) model with quartic contact interaction in the total density, 
\begin{equation}
 H =   \int  \mathrm{d}^3x \, \left[ -\Phi_{a}^\dag\frac{\nabla^2}{2m}\Phi_{a}  +  \frac{g}{2} \, \Phi_{a}^\dag\Phi_{b}^\dag\Phi_{b}\Phi_{a} \right] \,,
\label{eq:ONGPH}
\end{equation}
where we use units implying $\hbar=1$, space-time field arguments are suppressed, $m$ is the particle mass, and it is summed over the Bose fields, $a,b=1,2,3$, obeying standard commutators $[\Phi_{a}(\mathbf{x},t),\Phi_{b}^\dag(\mathbf{y},t)]=\delta_{ab}\delta(\mathbf{x}-\mathbf{y})$.
The gases are thus assumed to occupy the same space and be subject to identical inter- and intra-species contact interactions quantified by $g$.

Universal scaling of the $N$-component Bose gas \emph{at} the non-thermal fixed point can be described analytically in terms of a low-energy effective theory for the phase-angle excitations of the Bose fields $\Phi_{a}(\mathbf{x},t)=[{\rho_{a}^{(0)}+\delta\rho_{a}(\mathbf{x},t)}]^{1/2}$ $\exp\{i \, \delta\theta_{a}(\mathbf{x},t)\}$, on constant mean background phases $\theta_{a}^{(0)}=0$ and densities $\rho_{a}^{(0)}$.
After integrating out the density fluctuations $\delta\rho_{a}$, the linear modes of this effective model are given by the total phase $\sum_{a=1}^{N}\delta\theta_{a}$, with Bogoliubov dispersion $\omega_{\mathrm{B}}(\mathbf{k})  = \sqrt{\varepsilon_{\mathbf{k}} (\varepsilon_{\mathbf{k}} + 2 g \rho^{(0)} )}$, $\varepsilon_{\mathbf{k}}=\mathbf{k}^{2}/2m$, and $N-1$ gapless Goldstone excitations of the relative phases, e.g. $\delta\theta_{a}-\delta\theta_{1}$,
with free-particle dispersion $\omega_{\mathrm{G}}(\mathbf{k})=\varepsilon_{\mathbf{k}}$.
A scaling analysis of the kinetic equation $\partial_{t}f_{a}(\mathbf{k},t)= I[f](\mathbf{k},t)$ governing the momentum-space redistribution of the phase-angle excitations $f_{a}(\mathbf{k},t) = \langle\delta\theta_a(\mathbf{k},t) \delta\theta_a(-\mathbf{k},t)\rangle$ at the fixed point provides an analytical prediction  for $\alpha$ and $\beta$ \cite{Mikheev2018a.arXiv180710228M,Schmied:2018mte}.
Here, 
$I[f]$
is a quantum-Boltzmann-type collision integral involving scattering terms non-linear in the distributions $f_{a}$, arising from the non-linear couplings of the $\delta\theta_{a}$.
One obtains, for $N\to\infty$ as well as $N=1$, the values \cite{Mikheev2018a.arXiv180710228M,Schmied:2018mte}
\begin{align}
\beta=1/2,\qquad\alpha=\beta\,d=3/2\,,
\label{eq:betaalpha}
\end{align}
consistent with the results of  \cite{Orioli:2015dxa,Chantesana:2018qsb.PhysRevA.99.043620} for $N\to\infty$.
The relation between $\alpha$ and $\beta$ reflects the conservation of the $d$-dimensional integral $\int_{\mathbf{k}}f_{a}(\mathbf{k},t)$.
This particular fixed point has Gaussian character, i.e., in the limit $t\to\infty$, correlation functions factorize and the scaling of $f_{a}(\mathbf{k},t)$ implies the scaling of $n_{a}(\mathbf{k},t)$ as well as of higher-order correlators of the $\Phi_{a}$ \cite{Mikheev2018a.arXiv180710228M}.

%===============================================
\begin{figure}[t]
\includegraphics[width=0.97\columnwidth]{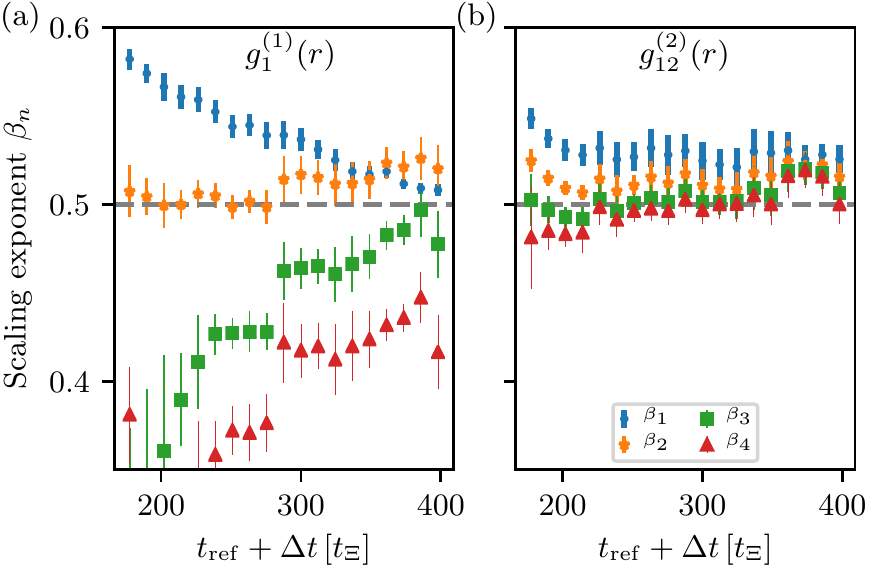}
\caption{Prescaling of position-space correlations.
(a)  Scaling exponents $\beta_{n}$ describing the time evolution of $k_{\Lambda,n}(t)\sim t^{-\beta_{n}}$ with $n =1,2,3,4$. The different
exponents are deduced from Taylor series coefficients $c_n(t) = c_n [k_{\Lambda,n} (t)]^n$ which are obtained by means of a fit of the first-order coherence function $g^{(1)}_{1}(r,t)$, shown in \Fig{RelativePhase12}a, at small distances $r$. The index $n$ marks the corresponding order of the Taylor series.
The jump of the exponents at $t_{\mathrm{ref}} + \Delta t  \approx 285\,t_{\Xi}$ results from a sign change of the fitted third and fourth order coefficients. This indicates that the shape of the scaling form is altered more significantly on large distances as compared to short distances as can already be expected from \Fig{RelativePhase12}a.    
 (b) Scaling exponents $\beta_{n}$ 
 deduced from an analogous Taylor series fit of $g^{(2)}_{12}(r,t)$ (\Fig{RelativePhase12}b).
Prescaling is quantitatively seen by the scaling exponents $\beta_n$ settling in to, within errors, equal stationary values for the lower orders of the fit.
While $g_{1}^{(1)}(r)$, up to order $r^{4}$ shows scaling violations, $g_{12}^{(2)}(r)$ already scales, to a good approximation, with the predicted exponent $\beta=1/2$ for $t_{\mathrm{ref}} + \Delta t \gtrsim250\,t_{\Xi}$.
For an individual fit, the $\beta_{n}$ result from averaging over times $[t_\mathrm{ref}, t_\mathrm{ref} + \Delta t]$ with $\Delta t = 146\,t_{\Xi}$.
The final data points shown are obtained by additionally averaging over a set of fits with different fit ranges. 
Errors are given by the corresponding standard deviation of the exponents of the set \cite{SupplMat}.
\label{fig:WindowFitAllkLi}
}
\end{figure}
%===============================================

Here, we numerically study the evolution of the system towards this fixed point, starting from a far-from-equilibrium initial condition at time $t_0$ given by large occupations of all fields,  $n_{0}\gg1$, constant up to some cutoff scale, i.e.~$n_{a}(\mathbf{k},t_{0})=n_{0}\Theta(k_\mathrm{q}-|\mathbf{k}|)$ \cite{SupplMat}.
The initial phase angles $\theta_{a}(\mathbf{k},t_{0})$ of the Bose fields $\Phi_{a}(\mathbf{k},t_{0})=\sqrt{n_{0}}\exp[i\theta_{a}(\mathbf{k},t_{0})]$ are chosen randomly on the circle and thus uncorrelated.
In practice, such an initial condition can be achieved by, e.g., a strong cooling quench or a transient instability \cite{Berges:2008wm,Chantesana:2018qsb.PhysRevA.99.043620}.
Note that already this initial state does not obey the full $U(3)$ symmetry but breaks it to $U(2)\simeq [S\!U(2)\times U(1)]/Z_{2}$ as does the evolving state.
The $U(3)$ symmetry of \eq{ONGPH} gives rise to conservation laws, consistent with the reduced $U(2)$ symmetry, which will be obeyed during prescaling \cite{Mikheev2018a.arXiv180710228M,SupplMat}. 
The evolution induced by such an extreme initial condition is characterized by transport of particles from $k\lesssim k_\mathrm{q}$ towards the infrared, while their energy is deposited by a few particles at higher momenta, $k>k_\mathrm{q}$. 
In this way the system, after a few collision times, shows universal scaling indicating the approach of a non-thermal fixed point \cite{Orioli:2015dxa,Berges:2015kfa,Chantesana:2018qsb.PhysRevA.99.043620}.

While the scaling behavior at a non-thermal fixed point is commonly extracted from momentum-space correlators, we find, however, that prescaling is more clearly seen in position-space correlations.
Based on momentum-space treatments of non-thermal fixed point scaling it is intuitive to study the first-order spatial coherence function $g^{(1)}_{a}(\mathbf{r},t)=\langle\Phi_{a}^{\dag}(\mathbf{x}+\mathbf{r}, t)\,\Phi_{a}(\mathbf{x},t)\rangle$, which is obtained as the Fourier transform of the occupation number $n_a(\mathbf{k})$. 
At large evolution times, close to the non-thermal fixed point, the coherence function is expected to be spherically symmetric and characterized by a universal function $f_\mathrm{s}( x)$ as $g_{a}^{(1)}(\mathbf{r},t) = f_\mathrm{s}( k_\Lambda (t) \, r)$,  $r=|\mathbf{r}|$.
The inverse coherence length scales as $k_{\Lambda}(t)\sim t^{-\beta}$.

The time evolution of the first-order coherence function is shown in \Fig{RelativePhase12}a.
We observe that the numerically extracted form clearly differs from a pure exponential, $g^{(1)}_{a}(\mathbf{r},t)\sim\exp\{-k_{\Lambda}(t)r\}$, which is predicted analytically within the leading approximation of a low-energy effective theory of non-thermal fixed points \cite{Mikheev2018a.arXiv180710228M}, 
with $k_{\Lambda}(t)$ being the inverse coherence length of the system at time $t$.
Instead, on top of an approximately exponential fall-off, the coherence function also shows oscillatory behavior in $r$. 
The oscillations indicate a structure developing in the system that causes excitations of the field to switch its sign over a distance on the order the inverse coherence length $k_\Lambda$, i.e.~the phase strongly varies on that characteristic scale.

We stress that, as the non-linear term in \eq{ONGPH} couples the total densities, it suppresses total-density fluctuations but not fluctuations of the local density differences between the components.
Hence, the spatial Goldstone excitations of the inter-component phase differences are predicted to become relevant.    
As the first-order coherence function is insensitive to the relative phases $\theta_{a}-\theta_{b}$, we additionally study
the second-order coherence function $g^{(2)}_{ab}(\mathbf{r},t)=\langle\Phi_{a}^{\dag}(\mathbf{x}+\mathbf{r},t)\,\Phi_{b}(\mathbf{x}+\mathbf{r},t)\,\Phi_{b}^{\dag}(\mathbf{x},t)\,\Phi_{a}(\mathbf{x},t)\rangle$, see \Fig{RelativePhase12}b, for $(a,b)=(1,2)$. 

A temporal scaling analysis of the numerically determined functions $g^{(1)}_{1}(\mathbf{r},t)$, $g^{(2)}_{12}(\mathbf{r},t)$ provides a direct way to extract the scaling exponent $\beta$ via the single scale $k_\Lambda(t)$. 
As long as, however, the fixed-point scaling is not yet fully developed, the time evolution of the correlations is not given by such a single scale.
To account for that we provide a general scheme for determining how the scaling behavior is being approached. 
In order to approximate the correlation functions, within a certain regime of $r$, without any restriction to a particular scaling form we expand them into a general Taylor series such that they take the form
$g^{(l)}(r,t) =c_0^{(l)} + \sum_{n=1}^{\infty} c_n^{(l)}(t) \, (r-r_0)^n$. 
Here, $r_0\geq0$ marks the expansion point, and $l=1,2$ denotes the two different types of correlators \cite{SupplMat}. 
The time-dependent coefficients of the series, dropping the $l$-index, are written as $c_n(t) = c_n [k_{\Lambda,n} (t)]^n$,   rescaling in time according to $k_{\Lambda,n} (t) \sim t^{-\beta_n}$.
In consequence, the coefficients of the expansion rescale as $c_n(t) \sim t^{-n \, \beta_n}$.
Each order of the expansion can be seen as a probe for the scaling of the correlations at a different distance $r$.
The corresponding scaling exponents can be written as $\beta_n(t) = \beta + \delta \beta_n(t)$. 
A particular order of the expansion shows scaling with the fixed-point exponent $\beta$ when $\delta \beta_n(t)$ becomes small and approximately constant in time. 
The system prescales when $\beta_n \approx \beta$ for at least one order $n$ of the expansion. 
The fixed point itself is, in a strict sense, only reached if the statement holds for all orders of the expansion.    

For our system we expect prescaling to emerge on short distances and to subsequently spread towards longer distances.
Therefore we truncate the expansion at the fourth order and extract the coefficients $c_n(t)$, with $n=1,2,3,4$, from a fit of the expansion to the data at various instances of time $t$. 
To focus on short-distance scaling properties of the system the fit is applied at distances  $5 \, \Xi  \lesssim r \ll \mathcal{L}$, with linear system size $\mathcal{L}$.
The lower bound of the fit range is used in order to not be affected by the non-universal short-distance thermal peak around zero distance.
Taking the negative of the logarithmic derivative of $c_{n}(t)$ with respect to $t$ and dividing by $n$ gives the scaling exponent $\beta_n$ at a particular instance in time. 
To reduce fluctuations of the locally in time extracted exponents we average the $\beta_n$ over a fixed time window.  
Taking into account possible fluctuations of the scaling exponents arising from the choice of the fit range  we furthermore average over different such ranges \cite{SupplMat}.
Performing the whole analysis procedure gives the scaling exponents $\beta_{n}$ shown in \Fig{WindowFitAllkLi}, for $n=1,...,4$, for both, $g_{1}^{(1)}$ and $g^{(2)}_{12}$.

The particular value $\beta_{n}\simeq0.5$ found, at late times, for the scaling of $k_{\Lambda,n}(t)$, for $n =1,2$, parameterizing $g_{1}^{(1)}$, and for $n=1,...,4$ in the case of $g^{(2)}_{12}$, is in good agreement with the analytically predicted value of $\beta=1/2$, cf.~\eq{betaalpha}
\cite{Orioli:2015dxa,Chantesana:2018qsb.PhysRevA.99.043620}.
Note that the finite size of the system does not lead to scaling beyond $t\simeq400\,t_{\Xi}$.

For $g^{(1)}_{1}$ we find that scaling in the higher orders of the expansion is not yet fully developed within our time window. 
This causes the scaling violations on larger distances observed in \Fig{RelativePhase12}.
The converging flow of the scaling exponents indicates the slow approach of a full scaling form.
In consequence, the system appears close to the non-thermal fixed point but is still away from it.

Comparing Figs.~\fig{WindowFitAllkLi}a and b we conclude that different correlators can enter the stage of prescaling on different time scales.
Therefore, establishing the full scaling function and the associated scaling exponents is observable-dependent.
This can also be intuitively concluded  from comparing Figs.~\fig{RelativePhase12}a and b. 
In general, we expect the scaling applying at the fixed point to first show up in correlators of observables that are most sensitive to the relevant degrees of freedom of the underlying universal behavior.
Hence, our results indicate that the fixed-point scaling of the model considered is dominated by relative-phase fluctuations, forming the Goldstone modes of the broken $U(3)$ symmetry \cite{Mikheev2018a.arXiv180710228M,SupplMat}.
Note that these excitations are much less energetically constrained than the sound-like excitations of the \emph{total} density, which are suppressed  by the interaction term in \eq{ONGPH} and associated with the overall $U(1)$ symmetry.
If $N$ is large, the relative-phase fluctuations, corresponding to spatial re-shuffling of the local density differences between the different components, will in general dominate the non-equilibrium evolution of the system, also of the single-component correlators $g_{1}^{(1)}$.
As $N=3$, however, is comparatively small, a clear difference in the scaling violations for $g_{1}^{(1)}$ and $g_{12}^{(2)}$ is seen.

We emphasize that the evolution during the stage of prescaling already obeys the conservation laws associated with the non-thermal fixed point.
Both, $n_a(\mathbf{k},t)$ and the Fourier transform of $g^{(2)}_{ab}(\mathbf{r},t)$ allow a scaling collapse according to \eq{NTFPscaling} with exponents $\alpha \simeq d\, \beta$.
This is consistent with number conservation reflecting the $U(3)$ symmetry of the Hamiltonian, as well as an emerging symmetry which ensures the invariance of $g^{(1)}_{a}(0,t)$ and $g^{(2)}_{ab}(0,t)$, respectively \cite{SupplMat}.

It is remarkable that the  $N=3$ prescaling exponents $\beta_{i}$ found for $g^{(1)}_{a}$ and $g^{(2)}_{ab}$ as shown in \Fig{WindowFitAllkLi} agree with the (for $N\to \infty$) analytically predicted value $\beta=1/2$ to a very good accuracy.
This suggests that the universality class of the model is independent of the number of components $N$, reflecting that the $U(N)$ symmetry is broken during prescaling to $U(N-1)$ and the dispersion of the dominating Goldstone relative-phase modes is independent of $N$.

A similar value for the scaling exponent $\beta$ has been found in recent experiments on a quasi one-dimensional three-component spinor Bose gas  \cite{Prufer:2018hto} which have motivated us to consider the $U(3)$ GP model.
In this experiment, additional spin-changing interactions and Zeeman shifts break the $U(3)$ symmetry, freezing out one of the relative-phase degrees of freedom at low $k$.
Nonetheless, given the experimental parameters, the measured momentum range is within a regime well described by the $U(3)$ model and prescaling is expected to be detectable.

Prescaling, observable in the relatively early evolution after a quench far from equilibrium, is expected to play an important role in universal scaling evolution and its accessibility in experiments with ultracold atomic gases. 
Furthermore, from a renormalization-group perspective and with respect to the given underlying symmetries we expect prescaling during the time evolution of various types of quantum many-body systems. 

\emph{Note added.}
After the completion of this work, Ref.~\cite{Mazeliauskas:2018yef} appeared, corroborating the prescaling predicted here.
 
%==============================================================================
%==============================================================================
\begin{acknowledgments}
The authors thank I.~Aliaga Sirvent, J.~Berges, K.~Boguslavski, R.~B\"uck\-er, I.~Chantesana, S.~Erne, F.~Essler, S.~Heupts, M.~Karl, P.~Kunkel, S.~Lannig, D.~Linnemann, A.~Mazeliauskas, J.~M.~Pawlowski, M.~K.~Oberthaler, 
A.~Pi{\~n}eiro Orioli, M.~Pr\"ufer, R.~F.~Rosa-Medina Pimentel, J.~Schmiedmayer, T.~Schr\"oder, H.~Strobel, and C.~Wetterich for discussions and collaboration. 
This work was supported by EU Horizon-2020 (AQuS, No. 640800; ERC Adv.~Grant EntangleGen, Project-ID 694561),  by DFG (SFB 1225 ISOQUANT), by DAAD (No.~57381316), and by Center for Quantum Dynamics, Heidelberg University.
C.-M.S.~thanks the Dodd-Walls Centre, University of Otago, NZ, for hospitality and support. 
T.G.~thanks the Erwin Schr\"odinger International Institute, Wien, for hospitality and support within their program \emph{Quantum Paths}.
\end{acknowledgments}
%==============================================================================

%==============================================================================
%==============================================================================
%==============================================================================
\onecolumngrid
\clearpage
%\newpage
%\begin{widetext}
\vspace{\columnsep}

\begin{appendix}
%\clearpage
%\widetext
\begin{center}
\textbf{\large Supplemental Material: Prescaling in a far-from-equilibrium Bose gas}
\label{SMpage}
%\textbf{APPENDIX}
\end{center}
\setcounter{equation}{0}
\setcounter{figure}{0}
\setcounter{table}{0}
\makeatletter

\renewcommand{\thesection}{S\arabic{section}}
\renewcommand{\theequation}{S\arabic{equation}}
\renewcommand{\thefigure}{S\arabic{figure}}

%%%%%%%%%%%%%%%%%%%%%%%%%%%%%%%%%%%%%%%%%%%%%%%%%%%%%%%%%%%%%
\vspace*{2ex}
In this supplemental material we provide further details of the numerical method and parameters used to simulate prescaling in position as well as momentum space, we discuss the implication of symmetries, summarize the signatures of prescaling in momentum space, and give more details of how we extract the scaling exponents.

%===================================================================
\section{Numerical method and parameters}
%===================================================================

All computations of time evolving correlation functions have been performed within the semi-classical truncated Wigner approximation (TWA) which is known to be well justified under the condition of high occupancies dominating the dynamics  \cite{Blakie2008a,Polkovnikov2010a}, as they are prevailing throughout the evolution described in our work.
For a justification of the semi-classical approximation in the framework of path integrals, cf., e.g., Ref.~\cite{Berges:2007ym}.
Since the universal scaling evolution towards a non-thermal fixed point is dominated by transport of particles in the infrared regime of highly occupied low momenta, the TWA is expected to provide a precise description of the dynamics.

The initial-state occupancy is chosen as $n_{0} \simeq 2350$, corresponding to a momentum cutoff $k_{\mathrm{q}} = 1.4\, k_{\Xi}$.
Here, $k_{\Xi} = \Xi^{-1}=[2 m g\, \rho^{(0)}]^{1/2}$ is a momentum scale set by the inverse healing length corresponding to the total particle density present in the system.
A spectral split-step algorithm is used to solve the coupled Gross-Pitaevskii equations derived from the Hamiltonian (2),
on a grid with $N_{\mathrm{g}} = 256^3$ points using periodic boundary conditions. 
The corresponding physical volume of our system is $V = {N}_{\mathrm{g}} \Xi^3$. 
The total particle number is $\mathcal{N} =\rho^{(0)}V= 6.7 \cdot 10^9$, i.e., we have $\mathcal{N}_a = 2.23 \cdot 10^9$ particles in each of the three components. 
The correlation functions are averaged over 144 trajectories.

%======================================================================================%
\section{Scaling at the non-thermal fixed point}
%=====================================================================================%

In this section we give a brief summary of the derivation of the universal scaling exponents $\alpha$ and $\beta$ predicted to characterize the scaling of the correlation functions at the non-thermal fixed point, i.e., in the scaling limit of ideally infinite evolution time. 
For more details, we refer to \cite{Mikheev2018a.arXiv180710228M}.
The derivation, in a path-integral language, makes use of the representation of the fluctuating Bose fields $\varphi_{a}$, $a=1,\dots,N$ (corresponding to the operators $\Phi_{a}$), in terms of the particle densities $\rho_{a}$ and phase angles $\theta_{a}$, 
\begin{equation}
\label{eq:Madelung}
{\varphi}_a (\mathbf{x},t) = \sqrt{\rho_a(\mathbf{x},t)}\, \exp\big\{\i \theta_a (\mathbf{x},t)\big\}\,.
\end{equation}
With these, the Lagrangian of the model (2), entering the action and thus the path integral reads, with total density $\rho=\sum_{a}\rho_{a}$, 
\begin{align}
\label{eq:Madelung_Lagr}
  \mathcal{L} 
  = -\sum_a \Bigg\lbrace  \rho_a \partial_t \theta_a 
   + \frac{1}{2m} \left[ \rho_a (\nabla \theta_a)^2  + (\nabla \!  \sqrt{\rho_a})^2 \right] \Bigg\rbrace 
   - \frac{g}{2} \rho^{2}\,.
\end{align}
The corresponding equations of motion include a continuity equation relating the density to the particle current $\mathbf{j}_{a}=\rho_{a}\partial_{\mathbf{x}}\theta_{a}/m$, and an equation for the phase $\theta_{a}$. 
In the limit of small fluctuations $\theta_{a}$ and $\delta\rho_{a}=\rho_{a}-\rho_{a}^{(0)}$ about the uniform ground-state densities $\rho_{a}^{(0)}= \langle \Phi_{a}^{\dagger}(x)\Phi_{a}(x) \rangle$, those equations reduce to the linearized equations of motion
\begin{align}
\label{app:dtdeltarhoa}
  \partial_t \theta_a &= \frac{1}{4 m \rho_a^{(0)}} \nabla^2 \delta \rho_a - g \sum_b \delta \rho_b\,, \quad
%  \\
  \partial_t \delta \rho_a = -\frac{\rho_a^{(0)}}{m} \nabla^2 \theta_a\,.
\end{align}  
In Fourier space, taking a further time derivative, they can be combined to the Bogoliubov-type matrix wave equation for the $\theta_{a}$,
\begin{equation}
\partial_t^2 \theta_a(\mathbf{k},t) + \frac{\mathbf{k}^2}{2m} \left(\frac{\mathbf{k}^2}{2m} \delta^{ab} + 2 g \rho^{(0)}_b \right) \theta_b (\mathbf{k},t) = 0\,,
\end{equation}
where Einstein's sum convention is implied. 
While, for $N=1$, we recover the Bogoliubov dispersion, for general $N$, diagonalization of the coefficient matrix yields the eigenfrequencies of $N-1$ Goldstone (G) and one Bogoliubov (B) mode, 
\begin{align}
  \omega_c(\mathbf{k}) 
  &\equiv\omega_{\mathrm{G}}(\mathbf{k}) 
  = \frac{\mathbf{k}^2}{2m}, \quad c=1,...,N-1\,,
%  \label{eq:GoldstoneFreq}
  \qquad
  \omega_N (\mathbf{k}) 
  \equiv\omega_{\mathrm{B}}(\mathbf{k})
  = \sqrt{\frac{\mathbf{k}^2}{2m} \left(\frac{\mathbf{k}^2}{2m} + 2 g \rho^{(0)} \right)}\,,
  \label{eq:GoldstoneBogHiggsFreq}
\end{align}
where $\rho^{(0)}=\sum_{a}\rho_{a}^{(0)}$ is the total condensate density. 
Note that the Goldstone theorem \cite{Goldstone:1961eq} predicts, due to the spontaneous breaking of the $U(N) \to U(N-1)$, $2N - 1$ gapless Goldstone modes. 
However, only $N$ of these modes, with frequencies \eq{GoldstoneBogHiggsFreq}, are independent due to the absence of Lorentz invariance and thus particle-hole symmetry \cite{Watanabe2012a.PhysRevLett.108.251602,Pekker2015aARCMP.6.269P}.
Hence, to take account of this fact and distinguish the modes, we only refer to the quadratic modes as Goldstone ones, whereas the linear one will be addressed as (hydrodynamic) Bogoliubov mode.

To describe the non-equilibrium transport of the quasiparticle excitations of the time-evolving correlation functions at the non-thermal fixed point, we derive a kinetic equation taking into account the interactions between the phase excitations introduced above.  
For this, we note that, at low energies, i.e., for $k \ll k_\Xi$, where 
$k_{\Xi} = [2 m \, \rho^{(0)} g]^{1/2}$
is a momentum scale set by the inverse healing length corresponding to the total density, the Bogoliubov mode contribution to the time derivative of fluctuations dominates, i.e.,
$\partial_t \delta \rho_a (\mathbf{k},t) \sim \omega_N (\mathbf{k},t) \delta \rho_a (\mathbf{k},t)$.
Then, according to \eqref{app:dtdeltarhoa}, below $k_\Xi$, the density fluctuations are decreasingly small compared to the mean density,
  ${\delta \rho_a(\mathbf{k})}/{\rho^{(0)}_a} 
  \sim ({|\mathbf{k}|}/{k_\Xi}) \theta_a(\mathbf{k}) 
  \ll \theta_a(\mathbf{k})$, 
  for
  $k \ll k_\Xi$,
such that we can approximate the dynamics by integrating out the $\delta \rho_a$ at quadratic order of the expansion of the Lagrangian \eq{Madelung_Lagr}. 
Applying the standard procedures we arrive at the effective action
 $ S_{\mathrm{eff}}=S_{\mathrm{eff,G}}+S_{\mathrm{eff,nG}}^{(3)}+S_{\mathrm{eff,nG}}^{(4)}$,
with Gaussian (quadratic) as well as three- and four-wave interaction parts 
\begin{align}
  &S_{\mathrm{eff,G}}[\theta] 
  = \int_{\mathbf{k}} 
  \frac{1}{2 } \Bigg\lbrace\frac{1}{g_{\mathrm{G}} 
  (\mathbf{k})} \left(\delta^{ab} - \frac{k_{\Xi,a} k_{\Xi,b}/k^2_{\Xi}}{1 + \mathbf{k}^2/2 k^2_{\Xi}} \right) 
  \partial_{t}\theta_a (\mathbf{k},t) \partial_{t}\theta_b (-\mathbf{k},t) 
 - \frac{\rho^{(0)} \mathbf{k}^2}{m} \theta_a(\mathbf{k},t) \theta_a (-\mathbf{k},t)\Bigg\rbrace,
 \label{eq:SeffGthetaa}
 \\
  &S_{\mathrm{eff,nG}}^{(3)}[\theta] 
  = \int_{\mathbf{k}\mathbf{k}'} 
  \frac{1}{N^{1/2}} \frac{1}{g_{\mathrm{G}} 
  (\mathbf{k})} \left(\delta^{ab} - \frac{k_{\Xi,a} k_{\Xi,b}/k^2_{\Xi}}{1 + \mathbf{k}^2/2 k^2_{\Xi}} \right) 
  \frac{k_{\Xi}}{k_{\Xi,b}}
  \frac{\mathbf{k}' ( \mathbf{k}' - \mathbf{k})}{2 m}
  \partial_{t}\theta_a (-\mathbf{k},t) \theta_b(\mathbf{k}',t) \theta_b(\mathbf{k} - \mathbf{k}',t)
  \Bigg\rbrace\,,
 \label{eq:SeffnG3thetaa}
 \\
  &S_{\mathrm{eff,nG}}^{(4)}[\theta] 
  = \int_{\mathbf{k}\mathbf{k}'\mathbf{k}''} 
  \frac{1}{2N} \frac{1}{g_{\mathrm{G}} 
  (\mathbf{k})} \left(
 \frac{ \delta^{ab} k_{\Xi}^{2}}{k_{\Xi,a}^{2}}
  - \frac{1}{1 + \mathbf{k}^2/2 k^2_{\Xi}} \right) 
  \frac{\mathbf{k}' ( \mathbf{k}' + \mathbf{k})}{2 m}
  \frac{\mathbf{k}'' ( \mathbf{k}'' - \mathbf{k})}{2 m}
  \theta_a(\mathbf{k}',t) \theta_a(-\mathbf{k} - \mathbf{k}',t)  
  \theta_b(\mathbf{k}'',t) \theta_b(\mathbf{k} - \mathbf{k}'',t)
  \Bigg\rbrace\,.
 \label{eq:SeffnG4thetaa}
\end{align}
Here, $\int_{\mathbf{k}} \equiv \int {\mathrm d^{d} k}/{(2 \pi)^{d}}$, and $k_{\Xi,a} = [2 m \, \rho_a^{(0)} g]^{1/2}$ is a momentum scale taking the form of the inverse healing length of a single component.
The Gaussian part has Luttinger-liquid form, with momentum-dependent coupling function
  $g_{\mathrm{G}}(\mathbf{k}) = {Ng \mathbf{k}^2}/({2 k^2_{\Xi}})$.
Note that the interaction terms \eq{SeffnG3thetaa} and \eq{SeffnG4thetaa} result from a basic three vertex between two phase-angle fields and one $\delta\rho_{a}$, arising in the expansion of the quadratic kinetic term in the GP model (2), while the cubic and quartic terms in the density fluctuations arising from the original non-linear term are being neglected.   
For simplicity, we consider, in the following, only the large-$N$ limit, while the above action allows to predict scaling exponents also for any finite $N\ge1$.
In this limit, we obtain \cite{Mikheev2018a.arXiv180710228M} 
\begin{align}
  S_{\mathrm{eff}}[\theta] 
  =& \int\limits_{\vec k} \frac{1}{2g_{\mathrm{G}} (\vec k)} \,\theta_a (\vec k, t)  \,  \left(-\partial_t^2 - (\vec k^2/2m)^2 \right) \theta_a(-\vec k, t)
  -\ \int\limits_{\lbrace \vec k_i \rbrace} 
  \frac{k_{\Xi,a}^2}{k_{\Xi}^2} \frac{N\vec k_1 \cdot \vec k_2}{2m\,g_{\mathrm{G}} (\vec k_3)} \, 
  \theta_a (\vec k_1, t)\, \theta_a (\vec k_2, t)\, \partial_t \theta_a (\vec k_3, t) \,\delta \Big(\sum_{i=1}^{3} \vec k_i\Big) 
  \nonumber\\
  &
  + \int\limits_{\lbrace \vec k_i \rbrace} 
  \frac{k_{\Xi,a}^2}{k_{\Xi}^2} \frac{N(\vec k_1 \cdot \vec k_2) \, (\vec k_3 \cdot \vec k_4)}{8m^2 \,g_{\mathrm{G}} (\vec k_1 - \vec k_2)} \, \theta_a(\vec k_1,t) \cdots \theta_a(\vec k_4,t)\, \delta \Big(\sum_{i=1}^{4} \vec k_i\Big)\,.
  \label{eq:Seff4}
\end{align} 

The kinetic description focuses on the evolution of equal-time two-point correlators, specifically on occupation number distributions of quasiparticles in momentum space, which due to symmetry, in our case are equal in all components $a$,
\begin{equation}
\label{eq:DefPhaseAngleCorr}
f_a(\mathbf{k},t) = \langle \theta_a(\mathbf{k},t) \theta_a(-\mathbf{k},t) \rangle.
\end{equation}
The kinetic equation for the time evolution governs the momentum spectrum $f_{\mathbf{k}}\equiv f(\mathbf{k},t)\equiv f_{a}(\mathbf{k},t)$ of phase excitations,
\begin{equation}
  \partial_{t}f(\mathbf{k},t)
  = I[f](\mathbf{k},t)\,,
  \label{eq:QKinEq}
  \qquad
  I[f](\mathbf{k},t)= I_{3}(\mathbf{k},t)+I_{4}(\mathbf{k},t)\,,
\end{equation}
where the scattering integral $I[f]$ is obtained, from the action \eq{Seff4}, with  \cite{Mikheev2018a.arXiv180710228M} 
\begin{align}
  I_3 (\mathbf{k},t) 
  \sim& \int_{\mathbf{p},\mathbf{q}} |T_{3}(\mathbf{k},\mathbf{p},\mathbf{q})|^2 \, 
  \Big[ (f_\mathbf{k} + 1) (f_\mathbf{p} + 1) f_\mathbf{q}  -  f_\mathbf{k} f_\mathbf{p} (f_\mathbf{q} + 1) \Big]\,,
  \label{eq:I_3}
  \\
  \label{eq:I_4}
  I_4 (\mathbf{k},t) 
  \sim& \int_{\mathbf{p},\mathbf{q},\mathbf{r}} |T_{4}(\mathbf{k},\mathbf{p},\mathbf{q},\mathbf{r})|^2 \,
  \Big[ (f_\mathbf{k} + 1) (f_\mathbf{p} + 1) f_\mathbf{q} f_\mathbf{r} -f_\mathbf{k} f_\mathbf{p} (f_\mathbf{q} + 1) (f_\mathbf{r} + 1) \Big] \,.
\end{align}
with the $T$-matrices expressed in terms of the coupling function, the momenta and the Goldstone dispersion $\omega(\mathbf{k})=k^{2}/2m$,
\begin{align}
\label{eq:T_3}
   |T_3(\mathbf{k},\mathbf{p},\mathbf{q})|^2 
   &= \left|\frac{(\mathbf{k} \cdot \mathbf{p})\, \omega(\mathbf{q})}{m\,g_{\mathrm{G}} (\mathbf{q})} + \text{perm}^{\text{s}}\right|^{2} 
   \frac{g_{\mathrm{G}} (\mathbf{k})\, g_{\mathrm{G}} (\mathbf{p})\, g_{\mathrm{G}} (\mathbf{q})}
   {8 \,\omega (\mathbf{k}) \, \omega (\mathbf{p}) \, \omega (\mathbf{q})}\,,
   \\
\label{eq:T_4}
  |T_4(\mathbf{k},\mathbf{p},\mathbf{q},\mathbf{r})|^2 
  &= \left|\frac{(\mathbf{k} \cdot \mathbf{p}) (\mathbf{q} \cdot \mathbf{r})}{2m^{2}\,g_{\mathrm{G}} (\mathbf{k} - \mathbf{p})} 
  + \text{perm}^{\text{s}}\right|^2   
  \frac{g_{\mathrm{G}} (\mathbf{k}) \cdots g_{\mathrm{G}} (\mathbf{r})}
  {2 \omega (\mathbf{k})  \cdots 2 \omega (\mathbf{r})}\,.
\end{align}

We now use the kinetic equations derived above to obtain a prediction for the scaling exponents $\alpha$ and $\beta$ at the non-thermal fixed point.
The fixed point is defined by the quasiparticle distribution obeying the scaling form
$f (\mathbf{k},t) = s^{\alpha/\beta} f(s \mathbf{k}, s^{-1/\beta}t)$ in space and time.
Since $g_{\mathrm{G}} (\mathbf{k}) = s^{-2} g_{\mathrm{G}} (s \mathbf{k})$ scales quadratically in momentum, the $T$-matrices scale as $|T_l(\mathbf{k}_{1},\dots,\mathbf{k}_{l};t)| = s^{-m_l} |T_l(s \mathbf{k}_{1},\dots,s \mathbf{k}_{l};s^{-1/\beta}t)|$, $l\in\{3,4\}$, with $m_3 = m_{4}=2$, resulting in the scaling of the scattering integrals as
$I_{l}[f](\mathbf{k},t) = s^{-\mu_{l}} I_{l}[f](s \mathbf{k},s^{-1/\beta} t)$,
with $\mu_l = 2 + (l-2)\, d  - (l  - 1)\,{\alpha}/{\beta}$.

The scaling distribution function $f$, at the fixed point, obeys the scaling form, and thus, for it to be a solution of the kinetic equation \eq{QKinEq} for a given $\mu=\mu_{l}$, the scaling exponents need to satisfy the relation $\alpha = 1 - \beta \mu$.
In addition, in the presence of global conservation laws for the integral $\int_{\mathbf k}f(\mathbf{k},t)$ (quasiparticle number), the scaling exponents are subject to the constraint $\alpha = \beta d$ such that both, $I_{3}$ and $I_{4}$ scale with $\mu_{l}\equiv\mu=2-d$. Collecting the above results, we obtain 
\begin{align}
\beta &= {1}/{2}\,,\qquad
\label{eq:betalargeN}
\alpha = {d}/{2}\,.
%\label{eq:alphalargeN}
\end{align}
Choosing $s=(t/t_\mathrm{ref})^{\beta}$ one obtains the scaling form $f (\mathbf{k},t) = (t/t_\mathrm{ref})^{\alpha} f_\mathrm{S}([t/t_\mathrm{ref}]^{\beta} \mathbf{k})$ of the distribution function. 
Moreover, by analysing the stationary scaling $f_\mathrm{S}(\mathbf{k})=s^{\kappa}f_\mathrm{S}(s\mathbf{k})$ of the scaling function $f_{\mathrm{S}}$ according to the kinetic equation, one derives the power-law behavior $f_{\mathrm{S}}(\mathbf{k})\sim k^{-d-1}$ at the non-thermal fixed point \cite{Mikheev2018a.arXiv180710228M} which eventually predicts the coherence function of each mode to evolve as $g^{(1)}_{a}(\mathbf{r},t) = \rho_{a}^{(0)}\exp\{-k_{\Lambda}(t) \, |\mathbf{r}|\}$, with $k_{\Lambda}(t)\sim t^{-\beta}$ in the scaling regime of large evolution times.

%======================================================================================%
\section{Scaling and conservation laws as seen in momentum space}
%=====================================================================================%
%
%===============================================
\begin{figure*}[t]
\includegraphics[width=0.88\textwidth]{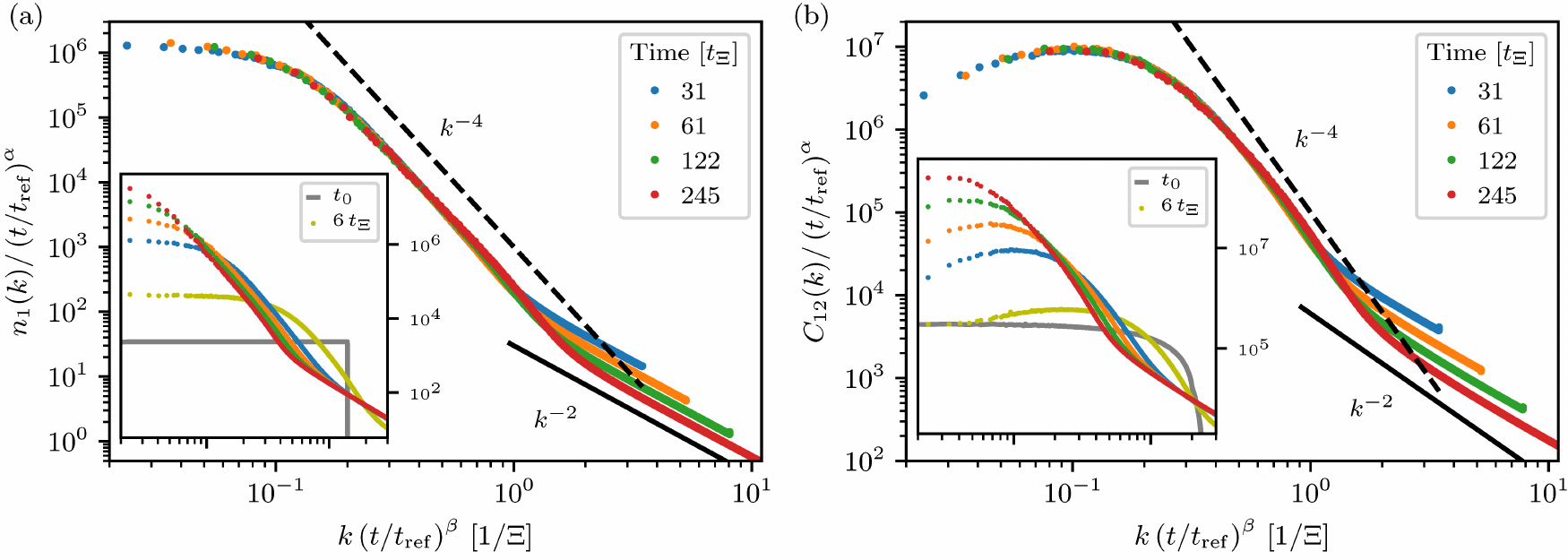}
\caption{\label{fig:NTFPEvolution} 
(a) Universal scaling of the occupation number $n_{1}(k)\equiv n_{1}(\mathbf{k},t) = \langle\Phi_{1}^{\dag}(\mathbf{k},t)\Phi_{1}(\mathbf{k},t)\rangle$. 
Inset: Evolution starting from a `box' momentum distribution $n_{1}(\mathbf{k},t_{0})=n_{0}\,\Theta(k_\mathrm{q}-|\mathbf{k}|)$, identical in all three components $a$ (grey line), with $n_{0}=(4\pi k_{q}^{3})^{-1}\rho^{(0)}$, $k_\mathrm{q}=1.4\,k_{\Xi}$, at five different times (colored dots). 
The collapse of the data to the universal scaling function $f_{\mathrm{S},1}(\mathbf{k})=n_{1}(\mathbf{k},t_\mathrm{ref})$, with reference time $t_\mathrm{ref}= 31\, t_{\Xi}$, 
shows the scaling in space and time.
Within the time window $t_{\mathrm{ref}}=200\,t_{\Xi} \leq t \leq 350\,t_{\Xi}$, we extract exponents $\alpha=1.62\pm0.37$, $\beta=0.53\pm0.09$, see  \Fig{WindowFit1}a.
At momenta  $k\gg k_{\Lambda}(t)$ we find a power-law fall-off of the distribution as $n_{1}(k)\sim k^{-\zeta}$ with $\zeta\simeq4$.
(b) Universal scaling dynamics of the correlator measuring the spatial fluctuations of the relative phases $C_{12}(k,t) = C_{12}(\mathbf{k},t) = \langle\lvert (\Phi_1^{\dagger} \Phi_2)(\mathbf{k},t)\rvert ^2\rangle $ for the same system. 
While no plateau prevails in the IR, a similar fall-off at higher momenta is seen as for $n_{a}(k)$, with only slightly modified power-law $C_{12}\sim k^{-4}$ in the scaling regime. 
 Within the same time window as stated in (a) we extract scaling exponents $\alpha=1.48\pm0.18$, $\beta=0.51\pm0.06$, see \Fig{WindowFit1}b. The scaling exponents characterizing the evolution of $C_{12}(k)$ are closer to the predicted fixed-point exponents  $\beta = 1/2$, $\alpha = 3/2$ than for $n_1(k)$.
}
\end{figure*}
%===========================================================================
%
%===============================================
\begin{figure}
\includegraphics[width=0.46\textwidth]{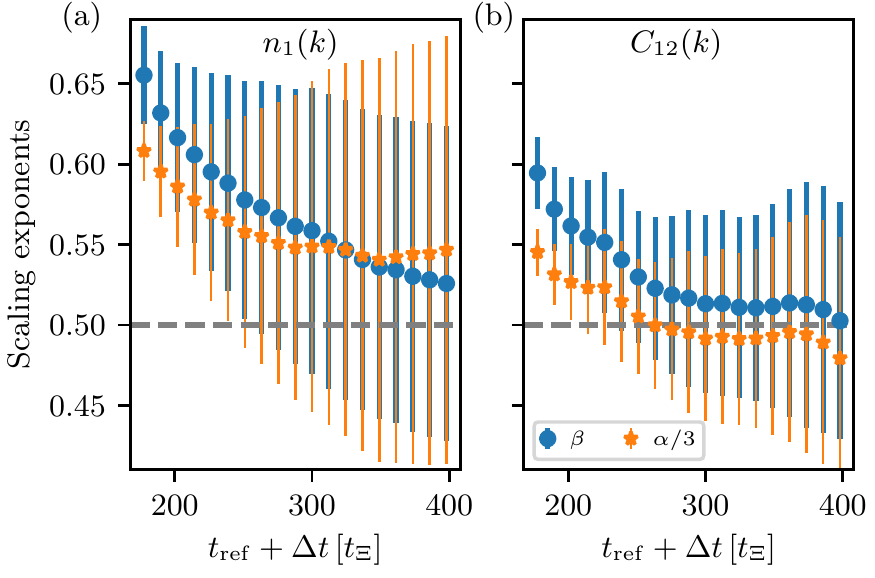}
\caption{(a) Scaling exponents $\alpha/3$ and $\beta$ obtained from least-square rescaling fits of the occupancy spectra $n_{1}(k)\equiv n_1(\mathbf{k},t)$ shown in \Fig{NTFPEvolution}a. 
The exponents correspond to the mean required to collapse the spectra within the time window $[t_\mathrm{ref}, t_\mathrm{ref} + \Delta t]$ with $\Delta t=146 \,t_{\Xi}$, and the momentum window $[k_\mathrm{min},k_\mathrm{max}]$, with $k_\mathrm{min}$ set by the lowest non-zero radial momentum at $t=t_\mathrm{ref}$ and $k_\mathrm{max}=0.45\,\Xi^{-1}$, such that the slight bend of $n_{1}(k)$ to a steeper power law is excluded. 
Error bars denote the least-square fit error.
 (b) Exponents extracted from the collapse of $C_{12}(k)\equiv C_{12}(\mathbf{k},t)$ shown in \Fig{NTFPEvolution}b, for the same time window.
While $n_{1}(k)$ still shows scaling violations, $C_{12}(k)$ exhibits approximate scaling at $t_\mathrm{ref} + \Delta t \gtrsim 300\,t_{\Xi}$ as the scaling exponents have settled in to stationary values. The scaling exponents are in good agreement with the analytically predicted values of $\beta =1/2$ and $\alpha=3/2$.
\label{fig:WindowFit1}
}
\end{figure}
%=============================================== 
%
In the main text, we emphasize that the evolution during the prescaling stage obeys the conservation laws associated with the  $U(3)$ symmetry which defines the universality class of the non-thermal fixed point. 
To demonstrate this conservation explicitly, we present, in the following, the scaling evolution of the correlations in momentum space, corresponding to the position-space observables displayed in the main text.
Note that, in previous work on non-thermal fixed points, as referenced in the main text, momentum-space correlations have been the usual object of studying scaling in space and time near a non-thermal fixed point, as they  clearly reveal the transport  associated with the evolution.

\Fig{NTFPEvolution}a shows that, within a range of low momenta, the evolution of the angle-averaged momentum distribution $n_{1}(k,t)=(4\pi)^{-1}\int d\Omega_{\mathbf{k}}n_{1}(\mathbf{k},t)$, with $n_{1}(\mathbf{k},t) = \langle\Phi_{1}^{\dag}(\mathbf{k},t)\Phi_{1}(\mathbf{k},t)\rangle$,  exhibits scaling in time $t$ and radial momentum $k=|\mathbf{k}|$ according to Eq.~(1) in the main text. 
While the inset shows snapshots of the evolution starting from the box initial distribution defined in the main text, the data shown in the main frame demonstrates the rather precise scaling collapse in momentum space, up to the scale on the order of the healing-length wave number $k_{\Xi}$ above which a near-thermal tail characterizes the higher-energetic particles.
Analogously, the coherence function $C_{12}(\mathbf{k},t)=\langle|(\Phi_{1}^{\dag}\Phi_{2})(\mathbf{k},t)|^{2}\rangle$, see \Fig{NTFPEvolution}b, shows a similar scaling collapse onto a scaling function revealing the same kind of universal scaling.

In  \Fig{WindowFit1}  we show that, as was found in position space, also the evolution of the momentum-space correlation functions is subject to scaling violations.
Both panels demonstrate that these violations prevail up to the maximum time $t\simeq400\,t_{\Xi}$, when finite-size effects have been found to become relevant.
During the late period, $t_{\mathrm{ref}}=200\,t_{\Xi} \leq t \leq 350\,t_{\Xi}$, one obtains the scaling exponents $\alpha=1.62\pm0.37$, $\beta=0.53\pm0.09$ for the scaling of $n_1(k)$, with a trend towards a smaller $\beta$, cf.~similar results found in \cite{Orioli:2015dxa}.
As already observed for the position-space correlations $g_{12}^{(2)}(r)$ discussed in the main text, the coherence function $C_{12}(k)$ shows much weaker scaling violations resulting in the scaling exponents $\alpha=1.48\pm0.18$ and $\beta=0.51\pm0.06$ within the time window $t_{\mathrm{ref}}=200\,t_{\Xi} \leq t \leq 350\,t_{\Xi}$, see \Fig{WindowFit1}b.
These findings are considerably closer to the exponents $\beta = 1/2$ and $\alpha = 3/2$  predicted for the fixed point,  see  Eq.~(3) and Refs.~cited in the main text. 
%===============================================
\begin{figure}[th!] 
\includegraphics[width=0.5\columnwidth]{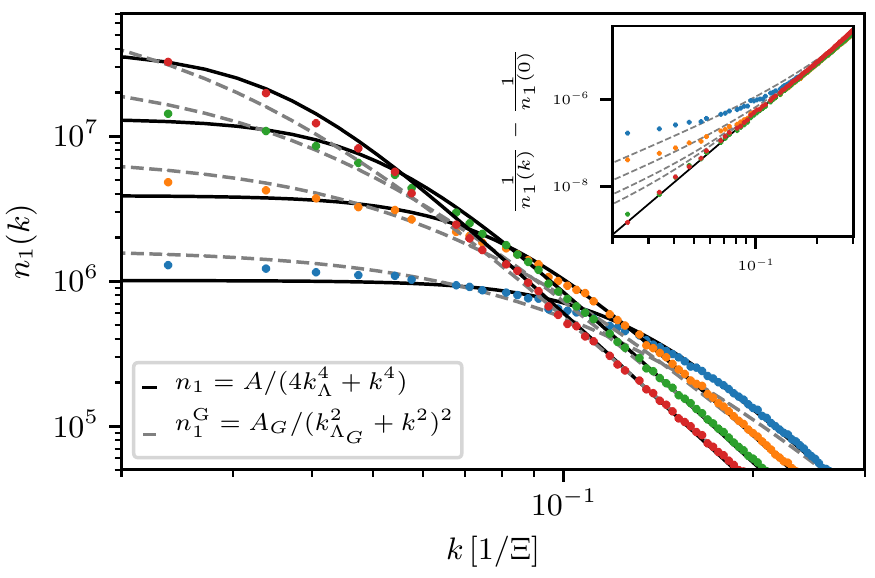}
\caption{Enlarged representation of the infrared prescaling evolution of the single-component occupation number $n_{1}(k)\equiv n_1(\mathbf{k},t)$, for the same evolution times as shown in \Fig{NTFPEvolution}a 
(same color coding). 
The solid black and dashed grey lines show the results obtained by fitting the corresponding scaling functions to the IR part of the distribution. 
The extracted parameters for the fit of $n_1$ are 
($A/C_3 k_\Lambda$, $k_\Lambda$, $t$) $=$
($0.82 \pm 0.02$, $0.088 \pm 0.001$, blue), 
($0.77 \pm 0.02$, $0.055 \pm 0.001$, orange), 
($0.76 \pm 0.02$, $0.036 \pm 0.001$, green), and 
($0.75 \pm 0.02$, $0.025 \pm 0.001$, red).
Analogously, the results for the fit of $n_G$ are 
($A_G/C_3 k_{\Lambda_G}$, $k_{\Lambda_G}$, $t$) $=$ 
($0.82 \pm 0.02$, $0.118 \pm 0.002$, blue), 
($0.76 \pm 0.02$, $0.070 \pm 0.001$, orange), 
($0.71 \pm 0.02$, $0.044 \pm 0.001$, green), and 
($0.69 \pm 0.02$, $0.030 \pm 0.001$, red). 
We emphasize that the thermal tail present in the regime of large momenta, see \Fig{NTFPEvolution}a, leads to an effective decrease of the extracted constant $C_3$ as compared to the analytical treatment, where a thermal tail is absent, such that the ratio $A/C_3 k_\Lambda$ differs from the analytically expected value of 1. 
Extrapolating the position-space correlation functions shown in Fig.~1a in the main text from distances larger than the thermal peak back to distance $r=0$ yields a factor of $\approx 0.8$ consistent with the ratios stated above from the fits.
At late times (green and red), the data is close to the scaling function $n_{1}(\mathbf{k},t)$, defined in \eq{NTFPscalingnk}, 
which corresponds to the first-order coherence function with exponential times cardinal-sine form, \eq{NTFPscalingg1}.
For all evolution times the data distinctly differs from the scaling function $n_{1}^{\mathrm{G}}(\mathbf{k},t)$ defined in \eq{app:NTFPscalingnkG} which corresponds to the purely exponential first-order coherence function \eq{app:NTFPscalingg1}.
This supports the observation of the presence of an oscillatory contribution in the first-order coherence function.
The fact that the quadratic term in the denominator gradually scales away is more clearly seen in the inset where we show $n_{1}(k)^{-1}-n_{1}(0)^{-1}$, with the respective extrapolated fit value inserted for $n_{1}(k=0)$ in order to be independent of possible deviations due to the buildup of a condensate in the zero mode. 
The solid black line in the inset corresponds to the fit of \eq{NTFPscalingnk} 
to the data for the latest time shown (red dots).
\label{fig:InverseOcc}
}
\end{figure}
%===============================================

In both cases, the scaling along the $k$-axis quantified by the exponent $\beta$ is coupled to the scaling along the vertical axis set by the exponent $\alpha$, related within errors by $\alpha= d\,\beta$, with $d=3$ being the spatial dimension of the system.
This coupled scaling leaves the total $d$-dimensional volume enclosed by the curves to a good approximation constant and thus, in each case, reflects a conserved symmetry during the prescaling evolution.

We emphasize that, due to the steep power law $n_{a}(k)\sim k^{-4}$ in the infrared (IR) scaling region, the particle number density
%, despite the multiplicative phase-space factor $\sum k^{2}$, 
is concentrated in this region of low momenta, where spatio-temporal scaling according to Eq.~(1) is seen.
In contrast, the energy density is concentrated in the high-momentum tail exhibiting semi-classical Rayleigh-Jeans distributions $n_{a}(k)\sim k^{-2}$.
Hence, during the scaling evolution, the scales at which particles and energy are concentrated, separate continuously further in time and thus, in the infrared (IR) region of momenta, this process is increasingly dominated by quasi-local particle number conservation only, see, e.g., Ref.~\cite{Chantesana:2018qsb.PhysRevA.99.043620}. 

This conservation law is due to the $U(3)$ symmetry of the model describing the dynamics of the system:
The initial state as well as the state during the ensuing prescaling evolution break the full $U(N)$ symmetry of the model to  $U(N-1)\simeq S\!U(N-1)\times U(1)/Z_{N-1}$, where the $S\!U(N-1)$ symmetry describes rotations of the resulting  state leaving the symmetry-broken mean-density vector invariant, and the $U(1)$ is the further total phase symmetry, also applying in the case of a vanishing mean-density vector. 
The $Z_{N-1}$ factor accounts for the non-isomorphic centers of $U(N-1)$ and $S\!U(N-1)$ but is irrelevant for the conservation laws considered here.
As a result, there is a remaining global $U(1)$ symmetry corresponding to the local conservation of total particle density in the $N$ components of the gas.
Our model does furthermore not allow for particle exchange between the different components, such that the particle numbers are actually conserved separately for each component.
Differently stated, taking the generators of the $U(3)$ group, one finds that each of the components carries a separate $U(1)$ symmetry which is conserved also in the broken state and its further evolution.
For the prescaling of $n_{a}(k,t)$, this implies the  relation $\alpha=d\,\beta$.

Note that our numerics exhibits the same type of conservation law and power-law behavior for the coherence function $C_{12}(k,t)$, see Figs.~\fig{NTFPEvolution}b and \fig{WindowFit1}b.
Here, the conservation law does not only refer to the conservation of the single-component and total particle numbers.
It furthermore reflects that the local density fluctuations are small.
This can be seen by considering the integral of $C_{12}(\mathbf{k})$ over all $\mathbf{k}$, which corresponds to the density-density correlation function,
\begin{align}
\int_{\mathbf{k}} C_{ab}(\mathbf{k},t) 
&=\ g_{ab}^{(2)}(0,t) 
= \langle \Phi_a^{\dagger} (\mathbf{x},t) \Phi_b (\mathbf{x},t) \Phi_b^{\dagger} (\mathbf{x},t) \Phi_a (\mathbf{x},t)  \rangle
 = \langle \delta \rho_a (\mathbf{x},t) \delta \rho_b (\mathbf{x},t) \rangle + \rho_a^{(0)} \rho_b^{(0)} + (\delta_{ab} - 1) \Lambda \rho_a^{(0)}\,,
\label{eq:g1220}
\end{align} 
where $\Lambda$ is a constant representing the momentum-space volume. 
As a result of the suppression of density fluctuations, ${\delta \rho_a(\mathbf{k})}/{\rho^{(0)}_a} 
  \sim {|\mathbf{k}|}/{k_\Xi} \theta_a(\mathbf{k})   \ll \theta_a(\mathbf{k})$ as compared to phase fluctuations $\theta_a(\mathbf{k})\sim\mathcal{O}(1)$ in the IR region of momenta  $k \ll k_\Xi$, cf.~\cite{Mikheev2018a.arXiv180710228M}, one may neglect the fluctuation contribution which implies an approximate conservation of the momentum integral of $C_{12}(\mathbf{k},t)$ in the scaling region, as seen in the relation $\alpha=d\,\beta$ between the scaling exponents also for $C_{12}(\mathbf{k},t)$.
Note that this symmetry is an emerging approximate symmetry which results during prescaling towards the non-thermal fixed point. 
This conservation law is expected to quantitatively improve the closer the system approaches the final fixed point.

The power-law fall-off, $n_{a}(k)\sim k^{-\kappa}$ for $k_{\Lambda}(t)\lesssim k \lesssim k_{\Xi}$, with $\zeta\simeq4$ is consistent with the analytically predicted exponent $\kappa=d+1$ in $d$ spatial dimensions, cf.~Ref.~\cite{Chantesana:2018qsb.PhysRevA.99.043620} as well as numerical results presented in \cite{Walz:2017ffj.PhysRevD.97.116011}. 
The analytic result is based on an analysis of the stationary fixed-point kinetic equation governing the spatio-temporally rescaled momentum distributions $n_{a}(\mathbf{k},t)$ \cite{Chantesana:2018qsb.PhysRevA.99.043620}, similar in character to a momentum scaling analysis in weak wave turbulence theory.

In the main text, we analyze quantitatively the position-space scaling function at comparatively short distances, larger than the extent of the thermal-peak and smaller than the scale of the first zero of the oscillating function.
In order to obtain more quantitative insight beyond the low-$r$, near-exponential fall-off, we in the following exemplarily compare two idealized limiting cases of the momentum-space scaling form.
Taking into account the extracted conservation law as well as the power-law fall-off, the momentum-space distribution $n_a(\mathbf{k},t)$ is approximately consistent with a scaling form given by 
\begin{equation}
n_{a}(\mathbf{k},t)=\frac{C_{3}\,k_{\Lambda}(t)}{4k_{\Lambda}^{4}(t)+|\mathbf{k}|^{4}}\,, 
\label{eq:NTFPscalingnk}
\end{equation} 
with normalization constant $C_{3}=8 \pi \rho^{(0)}_{a}$ and $k_{\Lambda}(t)\sim t^{-\beta}$, for $k\lesssim k_{\Xi}$, see \Fig{NTFPEvolution}a.
Fourier transforming the function \eq{NTFPscalingnk} gives an exponential $\times$ cardinal-sine form of the first-order coherence function,
\begin{equation}
 g_{a}^{(1)}(\mathbf{r},t) = \rho_{a}^{(0)}\exp\big(-k_{\Lambda}(t)\,|\mathbf{r}|\big)
 \,{\mathrm{sinc}\big(k_{\Lambda}(t)\,|\mathbf{r}|\big)}\,,
 \label{eq:NTFPscalingg1}
\end{equation}  
($\mathrm{sinc}({x})=\sin(x)/x$) with uniform particle density $\rho^{(0)}_{a}$.
An alternative functional form, which is also approximately compatible with the numerically determined momentum distribution, reads
\begin{equation}
n_{a}^{\mathrm{G}}(\mathbf{k},t)=\frac{C_{3}k_{\Lambda}(t)}{\left[k_{\Lambda}(t)^{2}+|\mathbf{k}|^{2}\right]^{2}}= \frac{C_{3}k_{\Lambda}(t)}{k_{\Lambda}(t)^4+ 2 k_\Lambda(t)^2 |\mathbf{k}|^2 +  |\mathbf{k}|^{4}}\,, 
\label{eq:app:NTFPscalingnkG}
\end{equation}
with normalization constant $C_{3}$ as above. 
In contrast to the functional form \eq{NTFPscalingnk}, a quadratic term  $\propto|\mathbf{k}|^{2}$ is added to the denominator. 
Note that the function \eq{app:NTFPscalingnkG} corresponds to the angle-averaged spatial first-order coherence function of the Bose field having the form of a pure exponential,
\begin{equation}
g^{(1)}_{a}(\mathbf{r},t) = \rho_{a}^{(0)}\exp\big(-k_{\Lambda}(t)|\mathbf{r}|\big)\,.
 \label{eq:app:NTFPscalingg1}
\end{equation}
A comparison of the numerical data and the above two different functional forms is shown in \Fig{InverseOcc}.
We find that, at late times, the data is close to the function \eq{NTFPscalingnk}.
In contrast, our data differs from the function \eq{app:NTFPscalingnkG} for all evolution times, which is in accordance with the observation of the presence of an oscillatory contribution in the first-order coherence function.

Due to the rescaling of the inverse coherence length as $k_\Lambda(t) \sim t^{-\beta}$, the quadratic term in the denominator sets in below a decreasing momentum scale during prescaling. 
This can be clearly seen in the inset, where, in the late-time scaling regime and within the region of momenta relevant in the finite-size system, a single power law prevails in the inverse of the momentum distribution after subtracting a constant.
Note, moreover, that the precise form of the function during the scaling evolution is in between both limiting cases, i.e., the quadratic term in the denominator is found to have a prefactor smaller than 2.
This results in a scaling function of the form of an exponential $\times$ cardinal-sine, with the argument $k_{\Lambda}(t)\,|\mathbf{r}|$ of the sine being multiplied with a different factor than that of the exponential.

The scaling violations in $n_{1}(k,t)$ corresponding to those discussed for $g_{1}^{(1)}(r,t)$ in the main text, are seen as a gradual change in time of the form of the distribution at low momenta, see \Fig{InverseOcc}. 
In contrast, scaling violations are again weaker for the case of the relative-phase fluctuations, as quantitatively seen in comparing Figs.~\fig{WindowFit1}a and b.

%======================================================================================%
\section{Numerical extraction of the scaling exponents}
%=====================================================================================%
In the main text we present a general scheme to extract the scaling behavior of the position-space correlators.
It is based on determining the coefficients of a Taylor series expansion by means of a fit. 
Here, we give details of the extraction procedure.

We approximate the angular-averaged correlation functions by a Taylor series of the form 
$g^{(l)}(\mathbf{r},t) =c^{(l)}_0 + \sum_{n=1}^{\infty} c^{(l)}_n(t) \, (r-r_0)^n$, where $r_0$ marks the expansion point of the series and $l=1,2$ refers to the two different correlation functions evaluated in this work.
Due to the presence of the non-universal short-distance thermal peak at very short distances, we choose the fit range to be limited by the lower bound $r_{\mathrm{l.b.}}$.
This lower bound furthermore is taken to be the expansion point of the Taylor series, $r_{\mathrm{l.b.}}= r_0$.
In order to stay consistent during the scaling evolution of the coefficients, $r_0$ is changed in time according to $r_0(t) \sim t^{\, \beta}$.
This is achieved by taking a fixed value of the correlation function at all considered instances of time and determining the distance associated with this value, i.e.~solving $g^{(l)}(r_0(t),t) =  g^{(l)}_{\mathrm{l.b.}}$ for $r_0(t)$.
Analogously we choose the fit to be limited by an upper bound $r_{\mathrm{u.b.}}$, which is defined by the relation $g^{(l)}(r_{\mathrm{u.b.}}(t),t) =  g^{(l)}_{\mathrm{u.b.}}$.
The upper bound of the fit was chosen in a range sensitive to the highest order of the expansion used to fit the numerical data.

To ensure equal quality of the fit on short distances, independent of the evolution time and of the choice of the lower bound of the fit, we optimize the residuals $ |g^{(l)}(r,t)/g^{(l)}_{\mathrm{fit}}(r,t)|$ of each fit at distances close to $r_0$ in a way that they symmetrically scatter around $1$ with a maximum deviation of $1\%$, i.e., we keep $0.99 \leq  |g^{(l)}(r,t)/g^{(l)}_{\mathrm{fit}}(r,t)| \leq 1.01$ in a range of distances $d_{\mathrm{min}}^{\,(l)}(t) \leq r(t) - r_0(t) \leq d_{\mathrm{max}}^{\,(l)}(t)$. 
The length of the interval $[ d_{\mathrm{min}}^{\,(l)}(t),\,d_{\mathrm{max}}^{\,(l)}(t) ]$ is given by  $d^{\,(l)} \cdot (t/t_{\mathrm{ref}})^{1/2}$, with reference time $t_{\mathrm{ref}} = 31 \, t_\Xi$. 
The parameter $ d^{\,(l)}$, characterizing the length of the interval at the reference time $t = t_{\mathrm{ref}}$, is chosen to be $d^{\,(1)} = 6 \, \Xi$ and $d^{\,(2)} = 4 \, \Xi$. 
At each instance of time, the minimal distance $d_{\mathrm{min}}^{\,(l)}(t) = r_{0,\mathrm{min}}^{\,(l)}(t) - r_0(t)$ is given by the difference between the lowest expansion point of all fit ranges used for a particular correlation function, $r_{0,\mathrm{min}}^{\,(l)}(t)$, and the lower bound of the current fit $r_0(t)$. 
Hence, $d_{\mathrm{min}}^{\,(l)}(t) = 0$ for the fit corresponding to the expansion point $r_{0,\mathrm{min}}^{\,(l)}(t)$, whereas $d_{\mathrm{min}}^{\,(l)}(t) < 0$ for all other fits
This means that, for fits that belong to the latter class, we ensure that it matches the data well even at distances below the expansion point $r_0(t)$. 
The maximal distance $d_{\mathrm{max}}^{\,(l)}(t)$ then directly follows from $d_{\mathrm{min}}^{\,(l)}(t)$ and from the fixed length of the interval $[d_{\mathrm{min}}^{\,(l)}(t),d_{\mathrm{max}}^{\,(l)}(t) ]$.

The exponents presented in Fig.~2 in the main text result from averaging over different fits with fit ranges of $0.55 \leq g^{(1)}_{\mathrm{l.b.}} \leq 0.65$,  $0.03 \leq g^{(1)}_{\mathrm{u.b.}} \leq 0.08$ for the first-order coherence function, and $0.25 \leq g^{(2)}_{\mathrm{l.b.}} \leq 0.40$,  $0.005 \leq g^{(2)}_{\mathrm{u.b.}} \leq 0.010$ for the second-order coherence function, respectively.
This defines $r_{0,\mathrm{min}}^{\,(1)}(t)$, at each instance of time and for all fits of the first-order coherence function, as the solution of $g^{(1)} (r_{0,\mathrm{min}}^{\,(1)}(t),t ) =  0.65$. 
Analogously, $r_{0,\mathrm{min}}^{\,(2)}(t)$ is determined by the solution of $g^{(2)}(r_{0,\mathrm{min}}^{\,(2)}(t),t) =  0.40$ for all fits of the second-order coherence function.

%===============================================

\end{appendix}

%==============================================================================
%==============================================================================
% Create the reference section using BibTeX:
\bibliographystyle{apsrev4-1}
%\bibliography{Bibliography/Master}

%merlin.mbs apsrev4-1.bst 2010-07-25 4.21a (PWD, AO, DPC) hacked
%Control: key (0)
%Control: author (72) initials jnrlst
%Control: editor formatted (1) identically to author
%Control: production of article title (-1) disabled
%Control: page (0) single
%Control: year (1) truncated
%Control: production of eprint (0) enabled
%
\onecolumngrid

%==============================================================================
%==============================================================================
\end{document}